\documentstyle[amssymb,aps,12pt]{revtex}

\begin{document}
\title{Influence of spin-orbit interactions on the cubic crystal-field
states of the $d^{4}$ system*}
\author{R.J. Radwanski}
\address{Center for Solid State Physics, S$^{nt}$Filip 5,31-150Krakow,Poland%
\\
Institute of Physics, Pedagogical University, 30-084 Krakow, Poland}
\author{Z. Ropka}
\address{Center for Solid State Physics, S$^{nt}$Filip 5,31-150Krakow,Poland.%
\\
e-mail: sfradwan@cyf-kr.edu.pl, http://www.css-physics.edu.pl\\
Original submission: 27.01.1997}
\maketitle

\begin{abstract}
It has been shown that for the highly-correlated $d^{4}$ electronic system
the spin-orbit interactions produce, even in case of the cubic crystal-field
interactions, a singlet ground state. Its magnetic moment grows rapidly with
the applied magnetic field approaching 4 $\mu _{B}$ for the $E_{g}$ state,
but only 3 $\mu _{B}$ for the $T_{2g}$ state. The applicability of the
present results to the Mn$^{3+}$ ion in LaMnO$_{3}$ is discussed.
\end{abstract}

\date{}

4 $d$ electrons of paramagnetic ions form a highly-correlated electron
system 3$d^{4}$ as their orbital and spin movements are correlated within
the incomplete outer shell. This Letter has been motivated by the recent
enormous increase of interest in La$_{1-x}$Ca$_{x}$MnO$_{3}$ compounds \cite%
{1,2,3}. Surely a large magnetoresistance effect, that enables wide
technical applications, vivifies interest to La$_{1-x}$Ca$_{x}$MnO$_{3}$
compounds. In the parent LaMnO$_{3}$ compound the manganese atoms occur in
the trivalent state anticipated from valences La$^{3+}$Mn$^{3+}$O$_{3}^{2-}$%
. This compound is an antiferromagnetic insulator with T$_{N}$ of 140 K and
the low-temperature magnetic moment of about 3.9 $\mu _{B}$ \cite{4}. The
presently discussed electronic structure\cite{1,2,3} of LaMnO$_{3}$
completely ignores the effect of spin-orbit interactions and the 5-fold spin
degeneracy of the $^{5}D$ term relevant to the 3$d^{4}$ electronic system.

The aim of the present Letter is to analyze the influence of spin-orbit
interactions on the localized states of the 3$d^{4}$ system produced by
crystal-field (CEF) interactions of the cubic symmetry. The 3$d^{4}$ system
is realized in paramagnetic ions like Mn$^{3+}$, Cr$^{2+}$, Ru$^{4+}$ (4$%
d^{4}$). The Mn$^{3+}$ state is realized in the antiferromagnetic insulator
LaMnO$_{3}$ that recently has got large interest as the parent compound of La%
$_{2/3}$Ca$_{1/3}$MnO$_{3}$ shows the Ca-induced ferromagnetism and the
enormous magnetoresistance effect at the vicinity of room temperature.

One can be surprised but the influence of the spin-orbit interactions has
not been systematically studied though CEF effect is already known by more
than 65 years, after the works of Bethe and Kramers at 1930. It is only
generally accepted that in 3$d$ ions spin-orbit interactions are much weaker
than CEF interactions - such the statement, however, has led to the
ignorance of spin-orbit interactions. Nowadays performed band-structure
calculations usually ignore spin-orbit interactions with a (wrong)
justification by the quenching-orbital-moment effect. The forgetting of the
spin-orbit interactions is, however, very dangerous. It will be shown in
this Letter that they produce a number of new physical phenomena like the
singlet ground state, for instance.

The physical situation of the 3$d^{n}$ system of a 3$d$-transition-metal ion
is taken to be accounted for by considering the single-ion Hamiltonian
containing the electron-electron d-d interactions H$_{el-el}$, the
crystal-field H$_{CEF}$, spin-orbit H$_{s-o}$ and Zeeman H$_{z}$
interactions:

$\ H_{d}=H_{el-el}+H_{CEF}+H_{s-o}+H_{z}$ \ \ \ \ \ \ \ \ \ \ \ \ \ \ \ \ \
\ \ \ \ \ \ \ \ \ \ \ \ \ \ \ \ \ \ \ \ \ \ \ \ \ \ \ \ \ \ \ \ \ \ \ \ \ \
(1)

The electron-electron and spin-orbit interactions are intra-atomic
interactions whereas crystal-field and Zeeman interactions account for
interactions of the unfilled 3$d$ shell with charge and spin surrounding.
These interactions are written in the decreasing-strength succession.

In a zero-order approximation the electron-electron correlations among the $%
d $ electrons are accounted for by phenomenological Hund's rules that yield
for the 3$d^{4}$ electron configuration the term $^{5}D$ with $S$ = 2 and $L$%
= 2 as the ground term. Under the action of the crystal field of the cubic
symmetry, the term $^{5}D$ splits \cite{5} into the orbital triplet $\Gamma
_{5}$, denoted also as $T_{2g}$, and the orbital doublet $\Gamma _{3}$ ($%
E_{g}$). Crystal field (CEF) effects of 3$d$ ions the reader is asked to
consult.

The $^{5}D$ term is 25-fold degenerated. The 5-fold degeneration occurs with
respect to the orbital degree of freedom and each orbital state contains the
5 spin-degree of freedom. This situation is accounted for if one considers
the single-ion Hamiltonian (1) written in the explicit form for the Hund's
rule $|LS\rangle $ term:

$H_{d}=B_{4}(O_{4}^{0}+5O_{4}^{4})+\lambda L\cdot S+\mu
_{^{B}}(L+g_{e}S)\cdot B_{ext}$ \ \ \ \ \ \ \ \ \ \ \ \ \ \ \ \ \ \ \ \ \ \
\ \ \ \ \ \ (2)

The first term is the cubic CEF Hamiltonian with the Stevens operators $%
O_{n}^{m}$ that depend on the orbital quantum numbers $L$ and $L_{z}$ and B$%
_{4}$ is CEF parameter. The second term is the spin-orbit interactions where 
$\lambda $ is the intra-atomic spin-orbit coupling. The last term accounts
for the influence of the magnetic field, the externally applied in the
present case. For simplicity the free-electron $g_{e}$ value is taken as 2.0
instead of 2.0023. The calculations of the many-electron states of the 3$%
d^{4}$ system have been performed \cite{6} by the diagonalization of a 25x25
matrix associated with the Hamiltonian (2) considered in the $%
|LSL_{z}S_{z}\rangle $ basis.

As a result of the diagonalization we obtain the energies of the 25 states
and the eigenvectors containing information e.g. about the magnetic
properties. These magnetic characteristics are computationally revealed
under the action of the external magnetic field. Due to the spin-orbit
coupling the states are no longer purely cubic orbital states $%
|L,L_{z}\rangle $ denoted as $|x^{2}-y^{2}\rangle ,$ $|3z^{2}-r^{2}\rangle ,$
$|xy\rangle ,$ $|zx\rangle ,$ $|yz\rangle $ states. In case of purely cubic
CEF interactions two first states form the orbital-doublet $E_{g}$ state
whereas the three last states form the orbital-triplet $T_{2g}$ state. These
states are shown in Fig. 1b for an exemplary value B$_{4}$=-10 K. (In
reality, B$_{4}$ amounts to 100 K - 300 K; here this small value of B$_{4}$
has been taken to show in the same scale the crystal-field and spin-orbit
interactions). In fig. 1c the effect of the spin-orbit interactions with $%
\lambda $ of +120 K ( +85 cm$^{-1}$, such the value is given for the Mn$%
^{3+} $ ion in ref. \cite{5} on p. 399). Inspecting Fig. 1 one can see that
spin-orbit interactions i) largely remove the degeneracies of the 10-fold $%
E_{g}$ state and of the 15-fold $T_{2g}$, state \ and yield ii) the singlet
ground state. This singlet ground state is produced always, i.e.
independently on the sign of the parameter B$_{4}$ as seen in Figs 1 and 2.
In Fig. 2 the same results are shown but for a positive value B$_{4}$ = +10
K. For a B$_{4}$%
\mbox{$>$}%
0 the orbital triplet $T_{2g}$ is the CEF ground state whereas for B$_{4}$%
\mbox{$<$}%
0 the orbital doublet $E_{g}$ is the ground state.

The reason for the formation of the singlet ground state is very simple. It
is, somehow, the manifestation of the 3$^{rd}$ Hund rule, that becomes valid
in case of the strong spin-orbit coupling (the situation realized in
rare-earth ions). The strong spin-orbit coupling is realized in Figs 1 and 2
by weakening of CEF interactions (going from left to right). The 3$^{rd}$
Hund rule says that the ground multiplet of the ($L$=2, $S$=2) term is $%
^{5}D_{0}$ with $J$=$L$-$S$=0 as the incomplete 3$d$ shell is less than half
filled. The $^{5}D_{0}$ multiplet is a singlet. Such the singlet $^{5}D_{0}$
multiplet occurs for the $d^{4}$ system where $\lambda $%
\mbox{$>$}%
0. The formation of the localized singlet ground state by the spin-orbit
interactions in the $d^{4}$ system, even in case of the doublet $E_{g}$
state, is a new and extremely surprising result owing to a general
conviction that the orbital-doublet state ($E_{g}$) is unaffected by the
spin-orbit coupling \cite{7}. It comes from this Letter that this is true
for the $^{2}D$ term applicable to the $d^{l}$ and $d^{9}$ system but not
for the $d^{4}$ and $d^{6}$ systems where $S$ = 2. This effect has been
signalized by Abragam and Bleaney \cite{5}, p. 435, who wrote that the $%
E_{g} $ state can be split by second-order effects, in the perturbation
method, of the s-o coupling. Now it has been proved by direct calculations
that these second-order effects have to be taken into account from the
beginning. For comparison of the effect of the spin-orbit coupling on the
cubic CEF states as has been calculated by Abragam and Bleaney and in the
present Letter the reader is asked to compare Fig. 1c and Fig. 7.19b on
p.404 of ref. \cite{5}. It allows for revealing of the novelty of the
present calculations. As mentioned already the Figs 1 and 2 have been made
for small values of B$_{4}$ parameters. The CEF splitting amounts usually to 
$\thicksim $25000 K, the spin-orbit splitting of the $T_{2g}$ state - to 600
K (5$\lambda $). Then the splitting of the $E_{g}$ state amounts to 14 K
only.

It is worth noting that despite of the singlet nature of the ground state
its magnetic moment grows rapidly with applying field as is shown in Fig. 3.
In case of the $E_{g}$ CEF ground state this induced moment approaches the
saturation to an integer value of 4 $\mu _{B}$, Fig. 3, curve 1. Such the
value is somehow manifestation of the well-known quenching of the orbital
moment in 3$d$ paramagnetic ions as just this value one expects for 4
electrons with the spin-moment only. For the $T_{2g}$ state the moment
saturates at level of 3 $\mu _{B}$ as one can see from Fig. 3, curves 3 and
4. This strong influence is due to van Vleck term and occurs so easily
because of the presence of a large number of low-energy localized states
close to the singlet ground state.

In conclusion, it has been shown that the spin-orbit interactions produce
for the highly-correlated $d^{4}$ system, even in case of the cubic CEF
interactions, the fine electronic structure with a singlet ground state both
for the octahedral (B$_{4}$%
\mbox{$<$}%
0) and tetrahedral (B$_{4}$%
\mbox{$>$}%
0) sites. As spin-orbit interactions are intra-atomic interactions it means
that one always deals with the singlet ground state in case of cubic CEF
interactions. Despite of its singlet nature the calculations show that its
magnetic moment grows rapidly with the applied magnetic field approaching
for the $E_{g}$ state (B$_{4}$%
\mbox{$<$}
0) a value of 4.0 $\mu _{B}$. Such the state is expected \cite{8} to be
realized in LaMnO$_{3}$ for which B$_{4}$ is anticipated for -200 K (and the 
$i$ ratio of 0.05) from the measured $d-d$ separation of 2.0-2.2 eV \cite%
{2,3}. The negative sign of B$_{4}$ is also consistent with the octahedral
ligand surrounding of the Mn$^{3+}$ ion in LaMnO$_{3}$. From the value of
140 K for T$_{N}$ in LaMnO$_{3}$ a molecular field of about 100 T has been
inferred. The present calculations clearly prove that for the meaningful
discussion of properties of 3$d$ paramagnetic-ion compounds the spin-orbit
and CEF interactions have to be taken into account at the starting point of
a physical analysis.

* This paper has been submitted to Phys.Rev.Lett. on 27 January 1997 but has
been rejected despite of our argument that our electronic structure is
basically different from that published in Phys.Rev.Lett., Refs 1-3. A
pretty long and intensive discussion has led to a scientific bet for 1
million dollars between the authors and the Editors. The reader is asked to
see our later papers.

References

Figure Captions

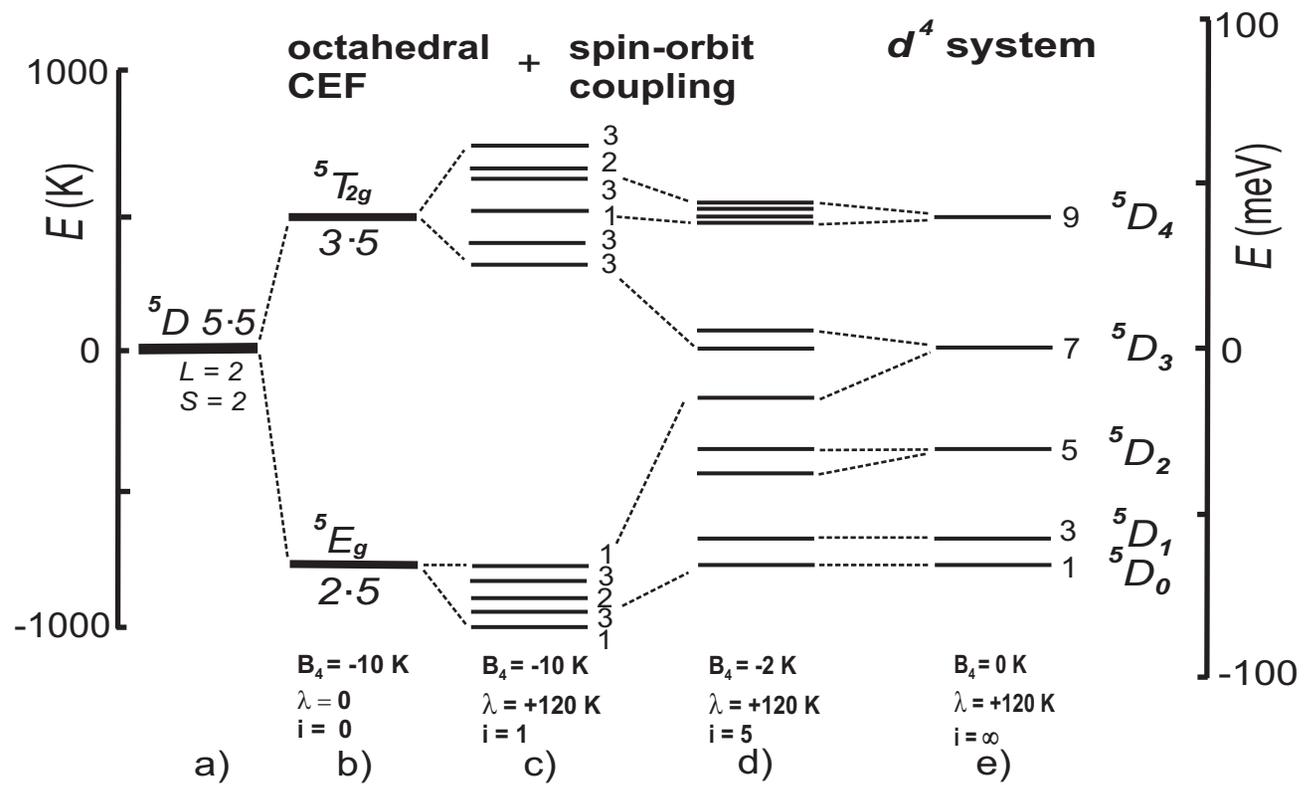
Fig. 1. Influence of spin-orbit interactions on the localized CEF states of
the 3$d^{4}$ system, that is expected to be realized in the Mn$^{3+}$, Cr$%
^{2+}$ and Ru$^{4+}$ (4$d^{4}$) ions, placed in the octahedral ligand field,
a) The $^{5}D$ term has 25-fold degeneracy; b) The energy level scheme of
the 3$d^{4}$ system under the action of CEF interactions of the cubic
symmetry, produced by octahedrally arranged O$^{2-}$ ligands, with B$_{4}$%
=-10 K, with the orbital-doublet $E_{g}$ ground state. All the levels would
have the internal 5-fold spin-degree of freedom that could be revealed in
the paramagnetic state by applying external fields; c) the effect of the
spin-orbit interactions with $\lambda $= +120 K on the localized states of
the 3$d^{4}$ system; the states are labelled with the degeneracy - a singlet
ground state should be noticed; d) the same as c but for smaller $|$B$_{4}|$%
, i.e. B$_{4}$=-2 K; e) the multiplet structure obtained for the extremely
large spin-orbit coupling realized here by the absence of CEF interactions; $%
i$ denotes the ratio of the overall spin-orbit splitting (=10$\lambda $) and
of the overall CEF splitting (=120 $|$B$_{4}|$). In real LaMnO$_{3}$, $i$
amounts to $\simeq $0.05, i.e. CEF\ interactions largely prevail the
spin-orbit coupling.

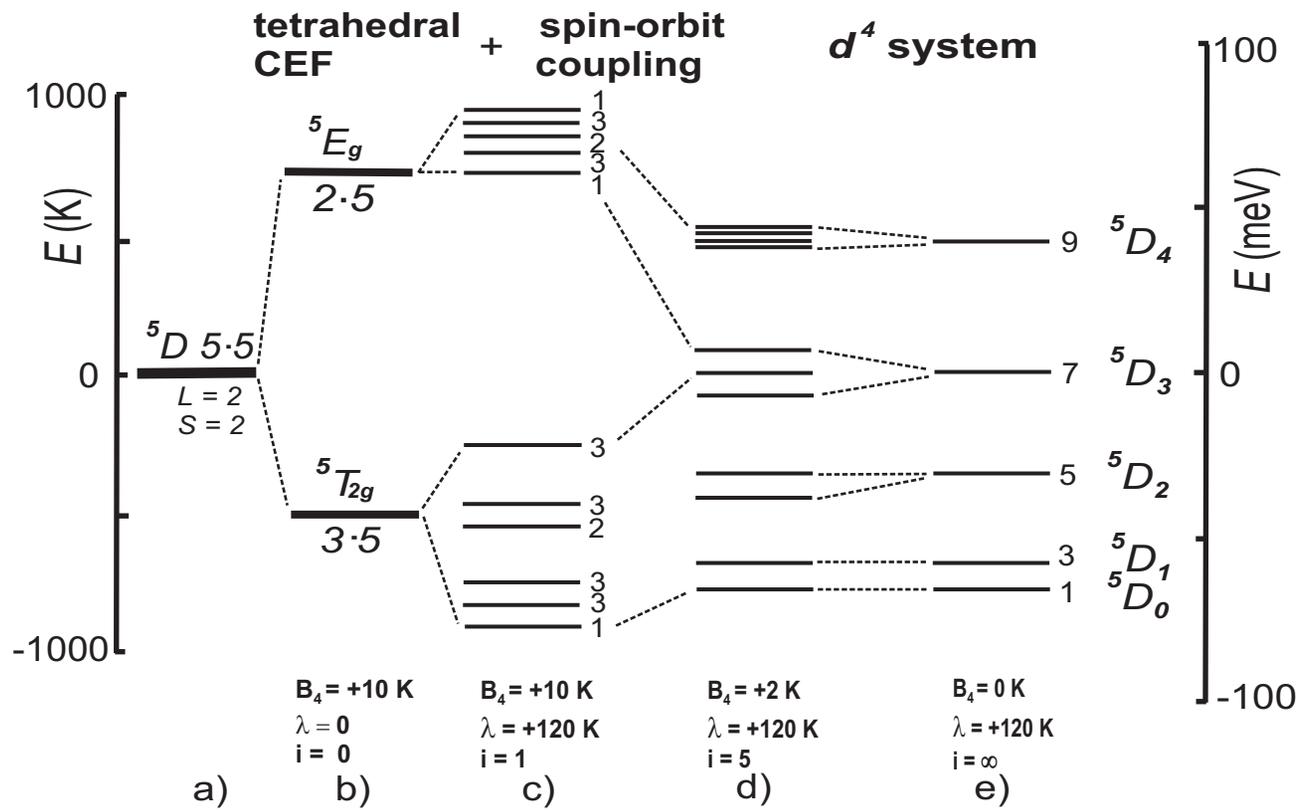
Fig. 2. The same as Fig. 1 but for the tetrahedral cubic CEF interactions,
i.e. B$_{4}$%
\mbox{$>$}%
0, that lead to the orbital-triplet $T_{2g}$ ground state.

Fig. 3. The magnetic-field induced magnetic moment of the 3$d^{4}$ system at
0 K with the $E_{g}$ (B$_{4}$%
\mbox{$<$}%
0, curves 1 and 2) and the $T_{2g}$ (B$_{4}$%
\mbox{$>$}%
0, curves 3 and 4) ground state calculated for $\lambda $=+120 K. The
dashed-point lines show results for $|$B$_{4}|$ = 100 K (curves 2, 4) and
the dashed line for $|$B$_{4}|$= 200 K (curves 1, 3). The induced-moment
somehow saturates to 2, 3 and 4 $\mu _{B}$ even in the case of magnetic
fields so large as 200 T. In LaMnO$_{3}$ the molecular field at 0 K is
anticipated for 100 T.


\begin{references}
\bibitem{6} the computer programme is available on the written request.
\end{references}
\end{document}